\documentclass[sigconf]{acmart}

\usepackage{dirtytalk}
\usepackage{graphicx}
\usepackage{subfigure}
\usepackage{multirow}
\usepackage{calc}

\usepackage{textcomp}
\usepackage{siunitx}
\usepackage{soul,xcolor}
\usepackage{gensymb}
\usepackage{enumitem}
\usepackage{booktabs}
\usepackage{acmart-taps}
\AtBeginDocument{%
  \providecommand\BibTeX{{%
    \normalfont B\kern-0.5em{\scshape i\kern-0.25em b}\kern-0.8em\TeX}}}

\copyrightyear{2025}
\acmYear{2025}
\setcopyright{cc}
\setcctype{by}
\acmConference[CHIWORK '25]{CHIWORK '25: Proceedings of the 4th Annual Symposium on Human-Computer Interaction for Work}{June 23--25, 2025}{Amsterdam, Netherlands}
\acmBooktitle{CHIWORK '25: Proceedings of the 4th Annual Symposium on Human-Computer Interaction for Work (CHIWORK '25), June 23--25, 2025, Amsterdam, Netherlands}
\acmPrice{}
\acmDOI{10.1145/3729176.3729202}
\acmISBN{979-8-4007-1384-2/25/06}

\begin{document}

\title[The Future of Work is Blended, Not Hybrid]{The Future of Work is Blended, Not Hybrid}

\author{Marios Constantinides}
\authornote{These authors contributed equally to this work.}
\orcid{0000-0003-1454-0641}
\affiliation{
    \institution{CYENS Centre of Excellence} \city{Nicosia} 
    \country{Cyprus}
}
\email{marios.constantinides@cyens.org.cy}

\author{Himanshu Verma}
\authornotemark[1]
\orcid{0000-0002-2494-1556}
\affiliation{
  \institution{Delft University of Technology}
  \city{Delft}
  \country{Netherlands}
}
\email{h.verma@tudelft.nl}

\author{Shadan Sadeghian}
\orcid{0000-0002-8590-656X}
\affiliation{
  \institution{University of Siegen}
  \city{Siegen}
  \country{Germany}
}
\email{shadan.sadeghian@uni-siegen.de}

\author{Abdallah El Ali}
\orcid{0000-0002-9954-4088}
\affiliation{%
  \institution{Centrum Wiskunde \& Informatica}
  \city{Amsterdam}
  \country{Netherlands}}
  \affiliation{%
    \institution{Utrecht University}
    \city{Utrecht}
    \country{Netherlands}}
\email{abdallah.el.ali@cwi.nl}

\renewcommand{\shortauthors}{Constantinides et al.}

\begin{abstract}
The way we work is no longer hybrid---it is blended with AI co-workers, automated decisions, and virtual presence reshaping human roles, agency, and expertise. We now work through AI, with our outputs shaped by invisible algorithms. AI’s infiltration into knowledge, creative, and service work is not just about automation, but concerns redistribution of agency, creativity, and control. How do we deal with physical and distributed AI-mediated workspaces? What happens when algorithms co-author reports, and draft our creative work? In this provocation, we argue that hybrid work is obsolete. Blended work is the future, not just in physical and virtual spaces but in how human effort and AI output become inseparable. We argue this shift demands urgent attention to AI-mediated work practices, work-life boundaries, physical-digital interactions, and AI transparency and accountability. The question is not whether we accept it, but whether we actively shape it before it shapes us.
\end{abstract}



\maketitle

\section{Introduction}

The way we work has fundamentally changed. The eight-hour workday, once the bedrock of industrial-era labor reforms, is eroding. Technology is now present in every aspect of work---blurring the already fragile boundary between professional and personal life. What used to be physical and co-located, human in its own right, is now algorithmic, dispersed, and mediated by Artificial Intelligence (AI). The workplace is no longer a ``place" \cite{Dourish2006}. It is an entanglement of tools, spaces, and AI systems that augment, assist, and sometimes even dictate our decisions. If the work structures we rely on today are artifacts of the Industrial Revolution, then flexibility was never meant to be part of the plan. It should come as no surprise that the shift toward hybrid, remote, and AI-integrated work is resulting in turbulence between organizations, employees, and employers. The challenge is not just in redefining productivity, well-being, and professional identity (cf., \cite{Kawakami2023sensingwellbeing,Das2024wellbeingphenotype})---it is in the fundamental struggle of letting go. Since when has quitting old habits been easy? We know change is needed, but are we truly willing to embrace it? Do we really want a future of work where AI is our colleague, our supervisor, our collaborator, our cooperator, our coordinator, and our creative co-author?

Technology has always shaped work, from the automation of manual labor to the AI-powered transformation of knowledge and creative work \cite{diamond1999guns, Basalla1989Evolution,Amershi2019,Muller2022, HumanAIInteraction:2024, SIG22FutureofWork}. In manufacturing, machines now outpace human workers in efficiency and accuracy \cite{Malik2024MAufacturing}. In healthcare, AI systems assist in diagnosis, treatment planning, and even remote patient monitoring, redefining the doctor-patient relationship \cite{meyer2021actithings, bhat2021infrastructuring}. Automated logistics and transportation are reducing human intervention in supply chains, while financial AI is making high-stakes trading decisions at speeds no human can match \cite{Cirqueira2021FraudDetection}. But what happens when AI does not just replace physical tasks, but begins replacing thinking? Generative AI systems such as ChatGPT, MidJourney, and DALL-E are not just tools---they are co-creators, altering how creative professionals, researchers, and writers produce knowledge \cite{hbr2023bgenerative,agrawal2023we, eloundou2023gpts, septiandri2024potential,Elagroud2024-LWM}, with strong potential for empowering workers and improving their wellbeing at word \cite{Das2024wellbeingphenotype}. In crowd work \cite{Kittur2013futurecrowdwork}, AI dynamically distributes tasks, fragmenting human labor into micro-contributions that are barely recognizable as jobs. Human-AI teams now execute knowledge-intensive work together, but with shifting definitions of expertise, authorship, and credit, as well bringing to question whether humans or AI are better alone or in combination \cite{Vaccaro2024,Schmutz2024}.

However, these transformations are likely not to be temporary disruptions but irreversible shifts that redefine the very fabric of employment, collaboration, and career progression (see Microsoft New Future of Work Report 2024~\cite{butler2024microsoft}). AI is not just another workplace tool---it is an active participant in decision-making, performance evaluation, and even the creative process \cite{Li2024}. If AI is blended into every aspect of work, what does it mean to be a skilled worker, a competent professional, or an accountable leader? In this provocation paper, we question our current notions of hybrid work \cite{ansah2023reflecting}, where we argue that blended work is the future, not just in physical and virtual spaces and places, but in the way human effort and AI output have become inseparable. We argue this shift demands urgent attention, where consider four aspects: (1) Designing for social relatedness in AI-mediated work can facilitate creativity, (2) demarcating and reinforcing work-life boundaries can improve worker well-being, (3) Rethinking the digital-physical continuum can enhance equity, accessibility, and inclusivity, and (4) Ensuring AI transparency and accountability can create fairer interactions across human-human and human-AI workflows. We are in a technological shift when it comes to creative and knowledge work practices, and it is no longer a question of whether we accept this shift, but rather how we can actively shape it before it shapes us. Below we dive into what Blended Experinces are, and why they should matter to the CHIWORK and HCI communities, as well as to design and research in general on current (knowledge and creartive) work practices. 

\section{What are Blended Work Experiences?}

Technology has always been integrated into work practices. Initially, it served as a means of augmenting human physical and cognitive resources—for example, by performing large calculations or storing and searching vast amounts of data. Over time, it evolved into a tool for supporting cooperative work. This trend accelerated significantly during the COVID-19 pandemic, which reshaped how we interact, work, and learn.
The increasing integration of these technologies has also had social implications. Technology is no longer just a tool for augmentation; it now determines when and through which medium "socializing" and teamwork occur in the workplace. While this shift has brought many benefits—such as enhancing opportunities for women and individuals with diverse abilities to participate in work collaboration—it has also introduced disadvantages. Hybrid communication has increasingly replaced in-person interactions, reducing opportunities for coworkers to connect beyond work-related topics and fulfill their fundamental psychological need for relatedness.

Since 2022, with the introduction of large language models (LLMs) such as ChatGPT and Bard, work practices have taken a new turn. The key question is no longer just about which tools we use or where work takes place, but rather \textit{who} performs the work and \textit{how} decisions about work processes are made. People now delegate tasks, converse, and collaborate with these AI systems. While this brings numerous benefits—such as increased efficiency and improved performance—it also raises important questions about work ownership, accountability, and the social perception of AI-assisted workers.
These technologies not only have the potential to replace human colleagues, affecting workplace relationships and impacting creative professionals \cite{Inie2023,Li2024} and artists \cite{Jiang2023}, but they can also impact social interactions by influencing how individuals are perceived by their peers. For instance, someone who relies on ChatGPT might be seen as less competent.

In this provocation paper, we specifically look into the individual and social implications of a blended work context on the work practices of knowledge and creative workers. First, we define our conception of ``blended'' in the context of AI-assisted knowledge and creative work. Similar to the notion of blended realities in virtual, augmented, and extended realities (e.g.,~\cite{robb2024blending}), where (physical/virtual) space and reality are superimposed with digital artifacts, information, and affordances, and where people's spatial and proxemic heterogeneity is remapped and blended to provide a seamless experience for collaborating members in mixed reality~\cite{wong2025spatial}. Blended work experiences manifest in the amalgamation of workers' tasks, outcomes, and responsibilities with the diverse possibilities realized by AI-enabled tools, processes, and environments. AI is no longer a \textit{means to an end}, a tool to serve one's utilitarian purposes; it is essentially diffused into work practices and outcomes in such a way that its presence and influence are rendered indistinguishable. Moreover, the role of AI in blended work is no longer about solving well-defined problems; rather, AI in blended work fundamentally transforms the horizon of possibilities, restructuring and rediscovering the socio-cognitive activities and transforming the creativity, accountability, and ownership that constitute knowledge work (see~\cite{verma_rethinking_2023, verma2021sensiblend}). This instrumental \textit{vs.} constitutive conception of AI in knowledge and creative work essentially distinguishes hybrid (and traditional) work from blended work experiences, where the former is concerned with using AI (and other digital tools) to support self-organization in relation to space, time, and colleagues, and does not fundamentally alter the constructs of autonomy, ownership, and accountability. As such, blended work also differs, albeit subtly, from other scholarly discourses at the intersection of AI and work, such as Human-AI Collaboration. The latter is primarily concerned with designing AI agents that can embody and express the qualities of an effective collaborator~\cite{wang2020from, gu2021lessons}, who can take the initiative from the ``human user'' and meaningfully engage in task (co-)management, negotiation, and process monitoring. In blended work, however, the aspects of human-AI collaboration may be present, but the focus is on the broader personal and societal implications that are fundamental, yet pervasive, to work and the workplace.

To clarify what we mean by blended work and how it differs from hybrid or traditional work, Tables \ref{tab:work_modes-Knowledge} and \ref{tab:work_modes_Creative} illustrate the differences in work practices for knowledge and creative workers across these various modes, along with their respective advantages and disadvantages. The examples in these tables are drawn from existing literature in HCI, CSCW, and AI and work (e.g., \cite{smids2020robots, anantrasirichai2022artificial, bamel2022managing} ), and illustrative insights from our own observations and professional experience. While not exhaustive, we aim to provide an understanding of the distinctions we draw conceptually and provide a comparative framework that illustrates how blended work diverges from prior models in both subtle and significant ways.
\smallskip

\begin{table*}[t]
    \centering
    \caption{Traditional, hybrid, and blended work modes for \emph{knowledge workers} along with their typical work routines, advantages, and disadvantages.}
    \scalebox{0.9}{
    \aptLtoX[graphic=no,type=html]{}{\setlist[itemize]{leftmargin=*}}
    
    \begin{tabular}{l|p{5cm}|p{7cm}}
        \hline
        \multicolumn{3}{c}{\textbf{Comparison of Knowlege Worker's Work Modes}} \\ \hline
        \textbf{Work Mode} & \textbf{Work Routine} & \textbf{Advantages \& Disadvantages} \\ 
        \hline
        \textbf{Traditional} & 
        \begin{itemize}
            \item Goes to the office in the morning
            \item Turns on PC and uses tools like Excel
            \item Talks to colleagues throughout the day
            \item Closes PC at 5 PM and goes home
        \end{itemize} 
        & 
        \textbf{Advantages:}  
        \begin{itemize}
            \item Clear work-life separation
            \item Face-to-face collaboration fosters strong workplace relationships
            \item Easier to receive immediate feedback and support
            \item Engages in informal conversations that enhance social well-being
            \item Full autonomy and ownership over designated tasks
        \end{itemize}
        \textbf{Disadvantages:}
        \begin{itemize}
            \item Commute time and costs
            \item Limited flexibility in work schedule
            \item Workplace distractions may reduce focus
            \item Restricted access to digital collaboration tools
        \end{itemize} 
        \\ 
        \hline
        \textbf{Hybrid} & 
        \begin{itemize}
            \item Works remotely from home
            \item Uses online tools like Zoom, Slack, and Google Docs
            \item Communicates with colleagues via video calls or chats
            \item Manages her own schedule but follows deadlines
        \end{itemize} 
        & 
        \textbf{Advantages:}  
        \begin{itemize}
            \item Increased flexibility and better work-life balance \cite{Duckert2023}
            \item No commute, saving time and costs \cite{obringer2021overlooked}
            \item Access to a variety of digital collaboration tools \cite{verma2023future}
            \item More autonomy in structuring the workday \cite{hocker2024healthy}
        \end{itemize}
        \textbf{Disadvantages:}
        \begin{itemize}
            \item Reduced face-to-face interactions may weaken social connections \cite{taser2022examination, sewell2015out}
            \item Potential feelings of isolation \cite{sewell2015out, taser2022examination}
            \item Harder to separate work and personal life \cite{verma2023future}
            \item More reliance on self-discipline to stay productive
        \end{itemize} 
        \\ 
        \hline
        \textbf{Blended (AI-Assisted Work)} & 
        \begin{itemize}
            \item Works remotely
            \item Collaborates with AI for scheduling, reporting, and content creation
            \item Uses AI-generated insights to assist decision-making
            \item Occasionally communicates with human colleagues
        \end{itemize} 
        & 
        \textbf{Advantages:}  
        \begin{itemize}
            \item High efficiency and productivity through AI automation \cite{hemmer2021human}
            \item Less time spent on repetitive tasks, allowing focus on creative work \cite{van2021hybrid}
            \item Greater flexibility in work hours \cite{ghosh2024impact}
            \item AI provides personalized recommendations and support \cite{dodeja2024towards}
        \end{itemize}
        \textbf{Disadvantages:}
        \begin{itemize}
            \item Reduced human interaction may impact social well-being \cite{smids2020robots}
            \item Risk of over-reliance on AI, leading to deskilling  \cite{rafner2022deskilling}
            \item Potential loss of job significance and ownership over work \cite{Sadeghian2024Soul}
            \item Colleagues may perceive AI users as less competent or replaceable \cite{Rae2024AIContentPerception}
        \end{itemize} 
        \\ 
        \hline
    \end{tabular}
    }
    
    \label{tab:work_modes-Knowledge}
\end{table*}

\begin{table*}[t]
    \centering
    \caption{Traditional, hybrid, and blended work modes for \emph{creative workers} along with their typical work routines, advantages, and disadvantages.}
    \captionsetup{justification=centering}
    \scalebox{0.9}{
    \aptLtoX[graphic=no,type=html]{}{\setlist[itemize]{leftmargin=*}}
    \begin{tabular}{l|p{5cm}|p{7cm}}
        \hline
        \multicolumn{3}{c}{\textbf{Comparison of Creative Worker's Work Modes}} \\ \hline
        \textbf{Work Mode} & \textbf{Work Routine} & \textbf{Advantages \& Disadvantages} \\ 
        \hline
        \textbf{Traditional} & 
        \begin{itemize}
            \item Goes to a creative studio or office in the morning
            \item Uses tools like Photoshop, Illustrator, or a writing pad
            \item Collaborates with teammates, brainstorming ideas in person
            \item Engages in creative discussions and informal chats
            \item Works on projects individually or in teams
            \item Leaves office at the end of the workday
        \end{itemize} 
        & 
        \textbf{Advantages:}  
        \begin{itemize}
            \item Rich face-to-face collaboration fosters idea generation
            \item Engages in spontaneous creative discussions
            \item Clear work-life separation
            \item Strong social connections with colleagues
        \end{itemize}
        \textbf{Disadvantages:}
        \begin{itemize}
            \item Commute time and expenses
            \item Distractions in a shared workspace
            \item Limited flexibility in working hours
            \item Dependence on in-person interactions for feedback
        \end{itemize} 
        \\ 
        \hline
        \textbf{Hybrid} & 
        \begin{itemize}
            \item Works remotely from home or a co-working space
            \item Uses digital tools (e.g., Figma, Miro, Slack) for collaboration
            \item Participates in online brainstorming sessions
            \item Has virtual check-ins with team members
            \item Creates content independently but submits for online review
        \end{itemize} 
        & 
        \textbf{Advantages:}  
        \begin{itemize}
            \item Increased flexibility and autonomy \cite{Duckert2023}
            \item No commute, saving time and costs \cite{obringer2021overlooked}
            \item Access to a wide range of online creative tools \cite{verma2023future}
            \item More control over work environment (fewer office distractions)
        \end{itemize}
        \textbf{Disadvantages:}
        \begin{itemize}
            \item Reduced spontaneous creative collaboration \cite{taser2022examination}
            \item Harder to receive immediate feedback \cite{bamel2022managing}
            \item Potential isolation, leading to less inspiration from colleagues \cite{taser2022examination, sewell2015out}
            \item More screen time, which can be mentally draining \cite{tang2020employee}
        \end{itemize} 
        \\ 
        \hline
        \textbf{Blended (AI-Assisted Work)} & 
        \begin{itemize}
            \item Works remotely and integrates AI tools (e.g., ChatGPT for ideation, MidJourney for visuals, Grammarly for editing)
            \item Uses AI for concept generation and enhancement
            \item Collaborates with AI and human colleagues in real-time
            \item Reviews and refines AI-generated outputs before finalizing work 
        \end{itemize} 
        & 
        \textbf{Advantages:}  
        \begin{itemize}
            \item AI speeds up creative processes (e.g., content drafts, design suggestions)\cite{han2024teams}
            \item More time for refining high-level ideas rather than routine tasks \cite{van2021hybrid}
            \item Flexibility to work from anywhere \cite{Duckert2023}
            \item AI assists in overcoming creative blocks \cite{Cetinic2022Understanding}
        \end{itemize}
        \textbf{Disadvantages:}
        \begin{itemize}
            \item Risk of over-reliance on AI, potentially reducing originality \cite{anantrasirichai2022artificial}
            \item Ethical concerns about AI-generated content (e.g., ownership, authenticity) \cite{lovato2024foregrounding}
            \item Less social engagement with colleagues \cite{nyholm2020can}
            \item Risk of AI-generated work being perceived as less valuable \cite{anantrasirichai2022artificial}
        \end{itemize} 
        \\ 
        \hline
    \end{tabular}
    }
    \label{tab:work_modes_Creative}
\end{table*}

  Before AI became pervasive in the form of widespread adoption of Large Language Models (LLMs), digital work revolved around humans using tools---video conferencing for meetings, document-sharing platforms for collaboration, and asynchronous messaging for communication. These technologies were designed as passive work facilitators that enable remote work while still keeping humans firmly in control of the workflow. However, in today's AI-infused workplace, an entirely new dynamic has been introduced. This new dynamic is characterized by AI that autonomously schedules meetings, drafts reports, and generates content, sometimes without explicit human initiation \cite{Holter2024}. AI now recommends, optimizes, and even mediates collaboration. AI systems are no longer just tools used by knowledge workers. Instead, they are active participants in the workplace, affecting accountability, authorship, and expertise~\cite{Sadeghian2022ArtificialColleague, smids2020robots,Dwivedi2023chatgptwroteit}.  
Recent research, particularly in the realm of AI's emerging role in cancer care and its relationship with physicians as a collaborator, has explored the \textit{instrumental} (i.e., means-to-an-end) vs. \textit{constitutive} (i.e., fundamentally (re-)shaping the working conditions as a member itself) conception of AI in the workplace~\cite{verma_rethinking_2023}.
This transition transforms work from human-centered to AI-mediated, making human responsibilities more fluid, ambiguous, and, at times, diminished. As AI co-workers, creative assistants, and productivity algorithms integrate into workplaces, they redefine social relationships between colleagues~\cite{nyholm2020can}, shift accountability for decisions \cite{Wieringa2020, martin2022excerpt, Sadeghian2024Soul}, and even alter employee motivation~\cite{Yakovenko2022AIWorkMotivation}. This is the essence of \emph{blended experiences}: a world where collaboration is not just hybrid (mixing remote and in-person work), but AI-augmented, adaptive, and partially autonomous.  
\smallskip

 It is tempting to conflate blended work with hybrid work, but they are fundamentally different. Hybrid arrangements describe the structural flexibility of work \cite{ansah2023reflecting, verma2021sensiblend}. In other words, they describe how individuals split their time between physical and virtual spaces, human and artificial interactions. Organizations implement hybrid work to improve self-organization, resource governance, and employee well-being~\cite{Kawakami2023sensingwellbeing} (e.g., flexible remote work policies). Blended experiences, on the other hand, are about how these hybrid arrangements manifest in practice. They describe the fluid mapping of work across digital, physical, and AI-mediated spaces. These experiences may include delegation of work to AI agents while retaining human oversight, embedding AI in creative workflows where human and machine-generated content are indistinguishable, or navigating an increasingly complex ecosystem of collaboration tools that autonomously assist, monitor, and intervene. To illustrate these points, consider the following example: a brainstorming session where an AI generates half of the ideas, a human curates them, and another AI summarizes the final concepts. \textbf{Blendedness is not just about where work happens---it is about who (or what) is performing it, how it is mediated, and how human roles evolve alongside AI.}
\smallskip

As AI becomes increasingly integrated into creative, managerial, and operational roles, our social contracts around work will be continuously renegotiated. The challenge is no longer just about how people navigate remote and in-person environments but rather about how they interact with AI-driven collaborators, AI-mediated tasks, and AI-influenced outcomes.  However, this raises a host of design, policy, and ethical questions~\cite{fjeld2020principled, constantinides2024implications, tahaei2023human, tahaei2023systematic, banovic2023being}. How do we foster serendipitous collaboration as well as deep focused work (cf., \cite{Xu2023returnoffice}) in blended environments where AI schedules, suggests, and filters interactions? How do we redraw the boundaries between work and personal life when AI tools persistently track productivity beyond office hours? How do we rethink accountability and authorship (cf., \cite{Elagroud2024-LWM,Schmidt2024, verma_rethinking_2023}) in HCI and AI-mediated work, where AI contributions may outweigh human effort? How do we ensure meaningful and effective AI disclosure (cf., \cite{Elali2024}) in an environment where human-AI collaboration or cooperation is indistinguishable?  

Blended experiences are no longer an experimental trend---they are the new normal. The question is not whether we will embrace them, but whether we will actively shape them or let them shape us. As such, we believe that our mission within CHIWORK and HCI in general is to de-couple or "de-blend" these experiences. This would enable us to more effectively grapple with the issues these blended realities are imposing on us: from authorship attribution, to accountability, to fair and just decision making, to decoupling spatial and temporal factors (cf., \cite{Ronott2024temporality}when we communicate with one another and with machines in this new AI-mediated future of creative and knowledge blended work. However, before we can de-blend, the HCI community needs to first shift the conversation from the opportunities and perils of hybrid work to blended work as our collective reality baseline, and only then operate within this new AI-mediated space and place.
\smallskip

\noindent\textbf{Why move beyond hybrid?} While hybrid work has become a dominant framing after the pandemic by offering flexibility across time and space, it largely assumes a human-centric model of collaboration where technologies act as facilitators. In contrast, blended work reflects a deeper entanglement between human and AI actors where decision-making, authorship, and task execution are increasingly shared. Hybrid arrangements describe where work happens, but blended work interrogates how work happens and who performs it. It acknowledges the increasing presence of AI as a co-worker, co-creator, and even supervisor. We argue that the HCI community must move beyond the spatial lens of hybrid work and confront the sociotechnical, ethical, and epistemological implications of AI-integrated labor.
\smallskip

\noindent\textbf{What does blended work contribute?} The conceptual move to blended work offers several contributions to HCI and work studies. First, it shifts the analytical focus from physical configurations of work to the relational and distributed agency of human and AI actors. Second, it emphasizes how creativity, decision-making, and accountability are negotiated in entangled human-AI workflows. Third, it surfaces new tensions (e.g., between transparency and opacity, spontaneity and automation, presence and mediation) that traditional hybrid framings overlook. Through blended work, we aim to offer new lens through which researchers and designers critically engage with sociotechnical realities of AI-mediated work.

\section{Blended Work in Practice: Four Considerations}
In this provocation, we focus on knowledge and creative workers; those whose work involves non-routine problem-solving, ideation, and symbolic manipulation, because they are disproportionately affected by the proliferation of AI tools. Their work is often performed autonomously, requires high cognitive input, and increasingly relies on digital systems for collaboration, content generation, and evaluation. As such, these workers are both early adopters and critical case studies for understanding how AI reconfigures the boundaries of expertise, authorship, and creative autonomy.

The shift toward blended work is about fundamentally rethinking how humans and AI collaborate across digital and physical spaces. While hybrid work policies focus on location flexibility, blended work is about how work itself is executed, shared, and mediated by AI. The question is not just how we structure work, but how we design for creativity, serendipity, well-being, accountability, and human-AI collaboration in a world where the boundaries between human effort and AI augmentation are becoming increasingly blurred. Below we present our four key considerations.

\subsection{Designing for Social Relatedness in AI-mediated Work Environments}
Serendipity has long been a catalyst for innovation, creativity, and collaboration in traditional workplaces. In physical office settings, chance encounters by the coffee machine, spontaneous brainstorming sessions, and informal hallway conversations often led to breakthrough ideas and unexpected solutions. These seemingly unstructured interactions played a critical role in fostering social bonds, creative problem-solving, and knowledge exchange, even if subsequent deep focused work creates the key output (cf., \cite{Xu2023returnoffice}). However, as work environments become increasingly AI-mediated and digitally structured, the natural flow of these moments is disappearing. In AI-blended work environments, meetings are scheduled by algorithms, team interactions are fragmented across multiple digital tools, and communication is often asynchronous, making spontaneous collaboration more difficult. Employees now rely on pre-scheduled Zoom calls, Slack threads, and AI-curated discussion topics, which prioritize efficiency over exploration. This structured approach reduces the organic cross-pollination of ideas, leading to what researchers describe as collaborative fatigue and social isolation in remote and hybrid work settings \cite{Gronbaek2021, Chiossi2024}. While digital collaboration tools aim to optimize work processes, they inadvertently create transactional interactions that fail to replicate the serendipitous encounters found in physical workspaces. As a result, creativity suffers, spontaneous problem-solving diminishes, and team cohesion weakens.  

To counteract this loss, we must rethink how AI and digital tools can enhance, rather than suppress, serendipity in blended work environments. Emerging solutions explore how technology can facilitate unplanned yet meaningful interactions, even in remote and hybrid settings. One avenue is through the use of Extended Reality (XR) technologies to create immersive and dynamic virtual spaces for social interactions. Systems like MirrorBlender \cite{Gronbaek2021} and Blended Whiteboard \cite{Gronbaek2024} have enabled users to engage in social interactions in virtual spaces (cf., \cite{Williamson2021,Ballendat2010}) even when geographically dispersed or solely within VR. These technologies provide a bridge between digital collaboration and spatial awareness, potentially making virtual meetings feel more like spontaneous, co-located experiences rather than pre-planned tasks.  Nevertheless, as most of these technologies aim to  mimic real-world interactions while being limited in their affordances, they create a gulf of execution, violating users' expectations. An alternative would be the benefit from unique possibilities of VR to augment human capabilities (e.g., being able to fly, or have unlimited memory) to create more engaging and stimulating forms of interaction \cite{Sadeghian2021VRSuperpower}.

Another avenue is the development of AI-enhanced social connection mechanisms. Instead of merely optimizing meetings for efficiency~\cite{aseniero2020meetcues, choi2021kairos}, AI systems could be designed to nudge employees toward spontaneous engagement by intelligently detecting potential discussion opportunities based on shared interests, past interactions, or complementary skill sets. For example, AI-driven platforms could suggest impromptu brainstorming sessions between colleagues working on similar problems or recommend cross-disciplinary team interactions that might not occur organically in a distributed work environment. By leveraging AI in this way, organizations can reintroduce elements of unpredictability and creative collision that fuel innovation.  

Furthermore, biosensing technology and affect-aware AI could enhance presence awareness in virtual and hybrid spaces, fostering a stronger sense of connection among remote collaborators. By integrating physiological and behavioral data, AI could adapt digital and virtual environments to create more immersive and socially rich interactions. Research suggests that subtle visualizations of users' engagement levels, emotional states, or cognitive focus (e.g., through real-time avatar adjustments or haptic feedback) could help recreate the nuanced non-verbal cues that are often missing in digital interactions \cite{Elali2023, Lee2022}. These biosignals and other physiological signals are also leveraged in designing virtual environments and agents that can adjust and respond appropriately and empathically based on the current emotional and cognitive state of the worker~\cite{gupta2024caevr, saffaryazdi2025empathetic}.
Empathic environments and agents are designed not only to enhance the employee experience, but also to make workplaces more inclusive by supporting individuals with social- and neurodiversity \cite{kim_workplace_2022}. 
LLM-powered conversational agents have also been afforded nuanced and rich personalities and behavioral traits to foster rich and organic human-AI interaction and user experience in diverse work-related contexts~\cite{kovacevic_chatbots_2024}.
Despite the empathic quality of AI-enabled and virtual agents facilitating novel and blended experiences in the workplace or more generally related to work, recent research on empathic conversational agents (CAs) has also provided evidence of how empathic agents can be used as tools for mass manipulation, social engineering, and persuasion~\cite{cuadra_illusion_2024, genc_persuasion_2025, pataranutaporn2023influencing}.
Although emerging research on the design and evaluation of empathetic agents and environments is providing new insights into how these technological developments affect workers' perceptions, experiences, and decision-making, it has also opened up new avenues of scholarly discourse that seek to critically reflect on the role of empathetic (AI-enabled) technologies in meaningfully and ethically supporting the future of work~\cite{empathich_2023, empathich_2024,Muller2024drinkingchai, nyholm2020can}. Nevertheless, these advances have the potential to transform virtual communication from a purely functional experience to one that is truly relational and engaging. 

The key challenge is not just increasing the quantity of digital interactions but fundamentally improving their quality. If AI-driven tools continue to prioritize structure over spontaneity, blended work environments risk becoming sterile and devoid of the creative friction that drives progress. Instead, the future of blended work must embrace AI-assisted serendipity, where technology facilitates unpredictable yet meaningful encounters, preserving the human element in innovation and collaboration.

\subsection{Designing for Job Meaningfulness and Wellbeing}
Blended work has reshaped not only where people work but also how deeply work permeates their personal lives. As physical offices, remote locations, and AI-mediated digital spaces merge, the once clear boundaries between professional and personal time are now fading away. AI-driven productivity tools, automated scheduling, and real-time performance tracking or self-tracking \cite{Gerdenitsch2023tracking} make it easier than ever to not only stay connected but also grow and improve. Unlike traditional work, where leaving the office signaled the end of the workday, today's AI-blended workflows extend beyond conventional time frames, pushing employees toward always-on availability through AI-generated reminders and optimized schedules. While these tools enhance coordination, they risk eroding the psychological separation between work and personal life, which is crucial for well-being and job satisfaction.  

One of the biggest challenges in blended work is boundary negotiation, that is, how employees define and maintain the separation between work and home~\cite{10.1145/3596671.3596672}. Cho et al. identified six types of boundary work that help remote workers structure their routines: spatial, temporal, psychological, sensory, social, and technological boundaries \cite{cho2022topophilia}. These range from physically separating workspaces at home to using digital tools that mark transitions between work and personal life, where proportions of what is ideal home office time may not be clear \cite{Colley2023, kun2024future}. However, AI-driven systems often ignore or undermine these boundaries and optimize for efficiency rather than human rhythms and cognitive needs. Another challenge is that AI has increasingly become a surveillance tool. AI-powered performance metrics, email monitoring, and engagement tracking have turned digital oversight into a source of anxiety and distrust. While organizations use these tools to measure efficiency, employees often experience them as invasive and stress-inducing. In the UK, one in two employees believes they are being monitored at work, while over two-thirds fear AI-driven surveillance could be used discriminatorily \cite{Cirqueira2021FraudDetection}. This creates a culture of digital presenteeism, where workers feel pressured to appear constantly engaged, even at the expense of creativity and deep work (which may arguably be more important than serendipitous encounters \cite{Xu2023returnoffice}). AI-driven productivity tracking further blurs the line between support and control, reinforcing unrealistic expectations and contributing to psychological distress.  

A deeper issue is that AI does not respect human rhythms---it optimizes for efficiency but not well-being. AI-generated scheduling, task delegation, and real-time notifications operate on an always-on logic, failing to account for burnout, cognitive fatigue, and personal work styles. For example, an AI may schedule a meeting based purely on availability data, disregarding a worker’s need for a mental break or time for focused work. Over time, this approach treats employees as mere nodes in a system by reducing flexibility and autonomy. To prevent AI from eroding work-life balance, AI systems must be designed to prioritize well-being alongside efficiency \cite{Kawakami2023sensingwellbeing,Das2024wellbeingphenotype}. AI-driven well-being assistants should not just push productivity but also encourage breaks, detect exhaustion, and promote deep work periods. Personalized work rhythms should give employees greater control over how and when AI assists them, allowing them to set boundaries, prioritize tasks, and manage interruptions in ways that align with their cognitive and psychological needs. Beyond individual solutions, organizations must implement AI governance policies to ensure transparency, autonomy, and ethical use of AI-driven monitoring. Workers should know what AI is tracking, why it is tracking it, and how that data is used (even if used for quantified self purposes \cite{Gerdenitsch2023tracking}). Without clear policies, blended work risks becoming a digital panopticon, where workers are expected to conform to AI-optimized expectations at the cost of well-being. The future of work must balance productivity with humanity, ensuring AI enhances work-life balance rather than eroding it.

At the same time, a critical lens on blended work must account for how AI mediation is experienced differently across hierarchical and labor divisions. Workers in lower organizational tiers or in precarious roles are more likely to encounter AI systems as instruments of \emph{surveillance, evaluation, and control} (cf., \cite{ball2022surveillance, sum2024s, constantinides2022good, das2023algorithmic}), rather than collaboration or augmentation. In contrast, managerial and creative roles may benefit from AI as a \emph{productivity partner} that offers relief from repetitive tasks. These asymmetries can exacerbate existing inequalities and entrench power dynamics in organizations. Any move toward blended work must therefore be accompanied by a focus on \emph{equity and empowerment} in a way that ensures AI systems support worker autonomy, dignity, and rights.

AI can also reshape the meaningfulness of work \cite{smids2020robots}, as has for example been shown by Kittur et al. \cite{Kittur2013futurecrowdwork} for job design within crowd work. Specifically, they showed that task design often fragments labor into micro-tasks that lack autonomy, skill variety, or a sense of purpose. More broadly, research shows that task significance, autonomy, and skill variety contribute to fulfilling work \cite{hackman1974job}. However, when AI takes over tasks that employees perceive as meaningful, it reduces their opportunities to contribute to their organization's goals. As a result, employees may experience a lack of autonomy and competence, have fewer opportunities to give or receive feedback from colleagues, and struggle to take credit for their creative work outcomes \cite{Sadeghian2024Soul, Sadeghian2022ArtificialColleague, Li2024}, even if creative attribution may not always be clear (see e.g., \cite{Sarkar2023creativity}).

\subsection{Rethinking Workspaces Across the Digital-Physical Continuum }
Blended work has redefined the workspace from a fixed location to an adaptive, interconnected environment where physical, digital, and AI-mediated interactions merge. Traditional offices were designed for co-located, face-to-face collaboration, but today’s workspaces must support fluid transitions between in-person, remote, and hybrid interactions. Existing tools like Zoom and Slack fail to replicate the rich, serendipitous, and dynamic nature of physical spaces, making work feel fragmented and transactional.  

To create workspaces that enhance creativity, deep focus, and social connection, we must move beyond basic digital replications and design intelligent, adaptive environments that blend the best of both physical and virtual worlds.  Next, we outline three directions about merging physical and digital workspaces, AI-driven adaptive environments, and XR for personalized workspaces. 

The first direction is to merge physical and digital workspaces. The future of hybrid workspaces lies in seamless spatial integration. Instead of treating virtual and in-person work as separate, organizations must create augmented meeting spaces where remote and on-site workers participate equally. Spatial computing, interactive digital walls, and AI-driven meeting facilitation can help bridge the gap, ensuring that virtual attendees are not passive participants but engaged collaborators~\cite{o2011blended}. In addition, existing barriers that impede seamless collaboration in mixed (physical and virtual) reality spaces, such as mismatch and heterogeneity in collaborating members' spaces, movements, and actions, would need to be addressed~\cite{wong2025spatial}. Without these adaptations, hybrid meetings risk reinforcing power imbalances, where in-person workers dominate discussions \cite{zhao2022mediated}. The second direction is about AI-driven adaptive environments. Workspaces should dynamically adjust based on context, task, and user needs. AI and IoT can adapt lighting, acoustics, and temperature to enhance concentration, comfort, and well-being. Research shows that workspace personalization improves focus and reduces stress \cite{zhao2022mediated}. AI can also analyze team dynamics, ensuring fair participation by amplifying remote voices when needed. The third direction is about XR for personalized workspaces. XR can transform workspaces into immersive environments that adapt in real time. A VR-based system could detect cognitive load and adjust distractions \cite{Chiossi2025}, creating focus-enhancing or socially interactive spaces as needed \cite{o2011blended}. Unlike current digital workspaces, which lack serendipity and embodied presence, XR offers spatial awareness and richer collaboration. However, accessibility and usability must remain priorities to ensure inclusivity.  
 
Blended work should not aim to digitally recreate the office—it should redefine what a workspace can be. Intelligent workspaces should actively enhance collaboration, creativity, and well-being, offering context-aware environments that support human needs.  The question is no longer where we work but how workspaces can intelligently support us. The future of blended work is not just about being connected—it is about designing environments that foster meaningful, seamless, inclusive, accessible, and equitable collaboration.
Despite the foreseeable opportunities to metamorphose and completely redefine workplace relationships and work practices, blended work, manifested through the reimagining of the physical-digital continuum, is not a panacea (or a universal remedy) for transforming work for the better. Designers, researchers, policy makers, and affected stakeholders and beneficiaries should be cautious and, more importantly, critically evaluate proposed developments in the physical-digital continuum. While this has the potential to make workplaces more inclusive, accessible, and meaningful (e.g., encouraging people to attend and interact with conference presenters), it can also be a tool for increased workplace surveillance, work-life imbalance, and harmful work praxis~\cite{ezra_systematic_2024, sum2024s, ball2022surveillance}.

\subsection{Ensuring AI Transparency and Accountability in Blended Work}
As AI takes on greater roles in decision-making, content generation, and workflow automation, a crucial challenge emerges: who is responsible when AI-mediated work goes wrong? AI is no longer just a passive tool but an active agent in the workplace, influencing everything from hiring recommendations and employee evaluations to report writing and marketing content. This shift raises fundamental concerns about accountability, credit, and transparency. When AI-generated decisions lead to mistakes or biases, the question arises: who should be held accountable—the human overseeing the AI, the organization deploying it, or the AI itself? Similarly, when AI significantly contributes to a work product, there is an unresolved achievement gap—should human workers take full credit, or does AI’s role necessitate new discussions about intellectual ownership? In essence, how can AI be continually designed responsibly \cite{OzmenGaribay07022023, nyholm2023responsibility} in this new future we are in? 

A major issue is the lack of clear AI disclosure in blended work environments. Many AI-driven tools function as black boxes, making decisions with little to no transparency about their underlying logic. Workers and decision-makers often interact with AI without realizing the extent of its influence, leading to ethical blind spots where AI-generated content is mistaken for human effort. While the European Commission’s AI Act (Article 50) mandates some level of AI disclosure, current policies remain vague \cite{LauxTrustworthyAI}, offering no concrete guidance on how AI transparency should be implemented in real-world workflows \cite{li2023european, Elali2024}. This ambiguity becomes especially critical considering the proliferating yet often opaque (secretive) usage of Large Language Models \cite{zhang2024secretuselargelanguage}, within and beyond hybrid work experiences. As such, developing truly effective AI disclosure strategies necessitates a comprehensive exploration of the user experience (UX) design space, examining disclosure mechanisms across diverse domains and interaction modalities \cite{Elali2024}. Moreover, as human-AI collaboration evolves in these blended environments, traditional disclosure paradigms face increased complexity. Beyond static, one-dimensional interactions, emerging collaborative and blended scenarios blur the boundaries between human and AI contributions \cite{hwang2024it80me20}, especially if accounting for temporal aspects in human-AI collaborations \cite{Ronott2024temporality}. Without stronger regulations and clearer AI accountability mechanisms, organizations risk deploying AI systems that lack interpretability, reinforce biases, or erode human oversight in decision-making.  

To address these challenges, AI in blended work must be designed with explicit transparency and accountability mechanisms \cite{OzmenGaribay07022023}. First, AI contributions should be clearly disclosed, ensuring that workers and decision-makers know when AI is involved in content creation, analysis, or decision-making. Second, human agency must remain central, allowing employees to override, question, or adjust AI-driven outputs rather than being passively guided by opaque systems. Finally, regulatory frameworks (within and beyond the EU AI Act and its limits \cite{Helberger21102023}) need to evolve beyond basic AI disclosure, addressing deeper concerns about bias, fairness, and authorship rights in AI-generated work (cf., creative professionals' worries about generative AI \cite{Inie2023}). As blended knowledge and creative work becomes an increasingly seamless fusion of human and AI effort \cite{Sarkar2023creativity}, the future of work must prioritize transparency, ethics, and accountability---ensuring that AI augments human (creative and knowledge) work without replacing oversight, authorship, or responsibility \cite{Li2024,Jiang2023}.

We believe the HCI community must resist the normalization of opaque AI co-authorship, especially in contexts where transparency, accountability, and critical interpretation are vital (e.g., scientific research, policy work). However, rather than wholly rejecting such practices, we call for a stance of \emph{critical acceptance}: one that demands that any delegation to AI, whether in writing, decision-making, or design, be accompanied by clear AI system disclosure (cf., \cite{Elali2024}), human oversight, and situated reflection, as has been done in the creation of this manuscript. Blended work should not disengage humans from authorship responsibilities, but compel us to redefine authorship as a shared but accountable process, even if attribution practices to AI systems and the systems themselves are rapidly evolving. To that end, we believe HCI researchers and designers are well-equipped to lead in drawing these boundaries proactively by shaping future best practices that ensure integrity, agency, and traceability, within and beyond the HCI discipline.

\section{Towards a Participatory Understanding and Design of Blended Work}
Blended work is no longer just a matter of hybrid schedules or digital collaboration---it is an entanglement of human, (Gen-)AI, and spatial interactions. The future of work will not be defined by where we work but by how seamlessly and intelligently these interactions are orchestrated. In this new paradigm, blendedness is not merely an operational shift but a fundamental reconfiguration of work itself, where co-located and remote humans, AI systems, and adaptive workspaces converge to create fluid, dynamic environments. To fully realize the potential of blended work, we must move beyond rigid distinctions between physical and digital, human and machine, presence and absence. Instead, we should strive for a model that fosters serendipitous interactions as well as focused individual deep work, reinforces work-life boundaries when needed, reimagines workspaces as adaptive environments, and ensures AI operates transparently and ethically. 

It is evident by now that contemporary discourse in AI development emphasizes the necessity of maintaining human-centric approaches while acknowledging the complexity of navigating diverse ethical perspectives across different stakeholder groups \cite{Jakesch2022ethicalai}. We argue this should apply to this new age of blended work. From the perspective of employers, organizations seeking to implement responsible AI practices must undertake comprehensive transformations of their existing workflows, considering current operations, emerging methodologies, and desired future states, as highlighted by Rakova et al. \cite{Rakova2021responsibleai}. This requires a multidisciplinary approach that not only enhances productivity but also safeguards human agency, collaboration, and creativity. To that end, we find that that Participatory AI methods (cf., \cite{Birhane2022,Inie2023}), in combination with value-sensitive design (VSD) methodologies \cite{Friedman1996}, are promising methods to support not only our understanding of AI technology (e.g., human-centered explainable AI \cite{Ehsan2020}), but also the complexities of and design for blended work practices through distilling human values onto AI for social good \cite{Umbrello2021valuedesign}. 

Participatory AI and VSD in our context offer clear methodological guidance -- they can help (re)frame how we understand the co-evolution of technology and labor, through first-hand accounts of talking to stakeholders. For example, participatory AI can be used to co-design generative tools for creative professionals (cf., \cite{Li2024}) by allowing users to define acceptable boundaries between what humans should focus on and what machine-generated outputs are better suited for. This allows collaborative negotiation of notions of authorship and addressing ethical concerns in situ. Likewise, VSD can surface human values such as autonomy, transparency, and accountability during the early phases of AI integration, whether these values are meant for co-authors, co-evaluators, or AI-enabled decision-making systems in the (future) workplace.

It is worth mentioning that while Self-Determination Theory \cite{ryan2017self} provides valuable insights into psychological needs, and has shown to be useful for understanding the needs of workers \cite{Gagne2022selfdetermination}, we believe participatory design methods in general \cite{Hansen2020}, and participatory AI in particular \cite{Birhane2022,Inie2023,Li2024}, are better suited for capturing the dynamic nature of our current human-AI collaboration in workplace settings. This allows us to examine work as it naturally unfolds with respect to the rapidly evolving pace of AI technology development, rather than through predetermined theoretical lenses or design principles. This is because blended work is not just about augmenting work---it is about rethinking what work means in an era where intelligence, presence, and labor are increasingly shared between humans and AI, and their respective configurations~\cite{Vaccaro2024}. It is continually evolving, and ongoing participatory AI insight gathering sessions are necessary for capturing the temporal nature of such human-AI collaborative practices (cf., \cite{Ronott2024temporality}). If blended work experiences are designed thoughtfully, it has the potential to transcend the limitations of both traditional and digital workspaces, creating environments that are not just connected, but truly blended—adaptive, equitable, and human-centered. The challenge now is not whether this shift will happen, but whether we will design it in a way that serves people, rather than reduces them to passive participants in an AI-optimized workflow.

Taken together, we believe a desirable future of blended work is not governed by humans adapting to AI, but rather a future where AI adapts to human values by \emph{amplifying agency, nurturing creativity, and supporting equitable participation}. This requires more than preserving existing human-centered values; it demands \emph{evolving} them in light of new forms of agency, collaboration, and interdependence. We envision blended work to go beyond the erosion of human roles, and serve as an opportunity to (re)build work systems that are transparent and co-created. Rather than framing blended work as a threat or a panacea, we see it as a design space; one where the future is not dictated solely by technology developments, but rather shaped by participatory, inclusive, and value-driven processes that place human life and values at the very center.

\section*{Positionality, Limitations, and AI Usage Disclosure Statement}

Recognizing the importance of author positionality is essential for transparently examining our perspectives on methodology and analysis~\cite{olteanu2023responsibleairesearchneeds,Frluckaj2022}. In this paper, we situate ourselves in three Western European countries (Cyprus, Netherlands and Germany) during the 21\textsuperscript{st} century, writing as authors primarily engaged in academic and industry research. We acknowledge that this paper addresses knowledge and creative work only, where work practices are broader and would likely require a different technological and policy lens (e.g., gig work \cite{Hsieh2023gigwork}). Our team comprises three males and one female from Asia and one from Southern Europe with diverse ethnic and religious backgrounds. Our combined expertise covers a range of areas, including human-computer interaction (HCI), ubiquitous computing, AI ethics, industrial design, and digital work. Our perspectives are shaped by our engagement with AI-mediated work environments, influencing both our critiques, choice of methods we advocate for, as well as the propositions in this paper.

In fact, this paper itself serves as an example of the blended and AI-mediated collaboration we discuss, and the need to disclose this usage. We used ChatGPT 4o large language model (LLM) as well as Claude (3.5 Sonnet) to refine clarity, structure arguments, and enhance readability. This illustrates how AI actively co-shapes knowledge production. However, we ascertain that the intellectual framing, the initial content created, the critical perspectives, and final interpretations remain our own. As AI increasingly entangles itself with human effort, we advocate for transparency in its influence, reflecting on how these tools shape authorship, accountability, and the evolving nature of work.

\balance
\bibliographystyle{ACM-Reference-Format}
\bibliography{main}

\end{document}